\documentclass[12pt]{article}
\usepackage{graphicx,epsfig}
\def\be{\begin{equation}}
\def\ee{\end{equation}}
\def\beq{\begin{equation}}
\def\eeq{\end{equation}}
\def\bea{\begin{eqnarray}}
\def\eea{\end{eqnarray}}
\def\t{$t~$}
\def\tbar{$\bar t~$}

\def\roots{\sqrt{s}}
\def\re{{\mathrm Re}}

\def\sigmat{\sigma(pp\rightarrow tW^-X)}  
\def\sigmatb{\sigma(pp\rightarrow \bar{t}W^+X)}
\newcommand\comment[1]{}

\begin{document}

 \begin{center}
 
 {\large \bf \boldmath CP violation in $tbW$ couplings at the LHC} 
 
 \bigskip
 
 Saurabh D. Rindani and Pankaj Sharma\footnote{
Present address: Korea Institute for Advanced Study, School of Physics,
207-43 Cheongryangri-dong,
Dongdaemun-gu,
Seoul 130-012, South Korea}
 
 \smallskip
 
 {\it Theoretical Physics Division, Physical Research Laboratory\\
 Navrangpura, Ahmedabad 380 009, India}
 
 \bigskip
 
 {\bf Abstract}
 
 \begin{quote}
 We study in a model-independent way anomalous CP-violating $tbW$
 effective couplings that might arise from new physics in the 
 processes $pp \to tW^- X$ and $pp \to \bar t W^+X$, followed by
 semileptonic decay of $t$ and $\bar t$. These processes have a dependence
 on effective $tbW$ couplings both in the production process as well as
 in the decay of the $t$ or $\bar t$. We propose several CP-violating asymmetries
 constructed out of variables in the two processes, including \t and
 \tbar polarization, and energy and azimuthal angles of the decay
 particles. We find that it is feasible to probe a certain 
 CP-violating combination
 of anomalous couplings at the per cent level at the LHC for centre-of-mass
 energy 14 TeV and an integrated luminosity of 10 fb$^{-1}$.
 
 \end{quote}
 
 \end{center}

\noindent
{\it 1. Introduction:}
The top quark, with a mass close to the scale of electroweak symmetry
breaking (EWSB), is generally believed to hold a key to the understanding of
the mechanism of symmetry breaking and responsible for masses of all
standard model (SM) particles. While the so-called Higgs mechanism is
the dominant scenario for EWSB, it has not yet been established
experimentally.  The currently operational 
Large Hadron Collider (LHC), which is proceeding full-steam with the
search for the Higgs boson,
can produce top quark-antiquark pairs in great profusion. 
It will shed light on the
details of the properties of the top quark, and hopefully, of
EWSB. Due to its large mass, the top quark decays before hadronization effects,
thus preserving polarization information in the decay products. The
study of the top-quark properties through its production and decay at the LHC 
would thus be of immense significance in arriving at a detailed
understanding of the one of the major mysteries of nature.

In addition to EWSB, another important phenomenon which lacks full
understanding is CP violation (CPV). Apart from having been seen
experimentally in mixing and decay of $K$ and $B$ mesons, CPV  
is also essential in understanding 
the observed baryon asymmetry of the universe. 
In the SM, the only source of CPV is the phase associated with the
Cabibbo-Kobayashi-Maskawa (CKM) inter-generational quark 
mixing matrix. However, the SM 
cannot adequately explain the baryon asymmetry \cite{AsymBar}. 
Extensions of the SM are needed to understand baryogenesis, and examples
of extensions which fit the bill are the two Higgs doublet models (THDM) and
minimal supersymmetric standard model (MSSM), which have additional
particles and newer mechanisms of CPV.
In fact, in these
models, it is the CP-violating scalar
couplings of the top quark 
which drive baryogenesis \cite{Model:baryogenesis}. Thus the study of
CPV, particularly  in the top-quark sector, could shed light on  primordial 
processes responsible for baryogenesis.

Apart from $t\bar t$ pair production,
which occurs dominantly through strong interaction,
single-top production, which necessarily proceeds via weak interaction,
also  has a large cross section at the LHC
\cite{Heinson:1996zm}-\cite{
AguilarSaavedra:2011ct}, and has already been seen
\cite{Chatrchyan:2011vp}. 
Single-top production will
therefore be able to give us information about the size and nature of
the weak-interaction
coupling of the top quark to the $b$ quark and the weak-interaction
gauge boson $W$, i.e., the $tbW$ coupling,
as well as the $tb$ (33) element of the CKM matrix.
Thus, for example, top polarization, which is negligible in the 
SM in top pair production because strong interactions conserve parity,
would be large in single-top production because of the chiral nature of
weak interactions \cite{Espriu:2002wx}.

In this work we investigate the possibility of probing 
putative CP-violating $tbW$ couplings in a model-independent way using 
single-top production in association with a $W$ boson. 
There are a number of proposals for the study of CPV in
top-pair production at lepton \cite{leptonic:CP},
photon \cite{photonic:CP} 
and hadron colliders \cite{hadronic:CP}. However, CPV in single-top production has
received little attention \cite{singletop:CP,singletop:FC}. The reason can be seen to be two-fold.
For one, unlike in $t\bar t$ production, where the final state is
self-conjugate, in single-top production, a process involving $t$ has to
be related to another involving $\bar t$, which is not straightforward.
The other reason is the low event rate for single-top production
expected at Tevatron. Single-top production at the LHC can be
substantial, and it would be worthwhile attempting to extract
information on CPV, even though it needs more elaborate
analysis than in the case of pair production.
Recent earlier work on CPV in single-top production at the LHC has been in
$tH^-$ associated production \cite{singletop:CP} or in the context of
flavor-changing top couplings \cite{singletop:FC}, but not so far in
$tW^-$ production.
This, therefore, is the first time that the possibility of
studying CP-violating $tbW$ couplings at the LHC is being explored.

There are three distinct mechanisms for single-top production, 
viz.,  a) the $t$-channel process $bq\rightarrow tq^\prime$, b) the
$s$-channel process $q\bar q^\prime\rightarrow t\bar{b}$ 
and c) the $tW$ associated production process $bg\rightarrow tW^-$
\cite{Tait:1999cf}. Although the process a) has the largest cross
section, the other processes also occur at a significant level. Since 
each process has a distinctive final state, it would be possible to
study each one of them separately. 

The process c), in which we are interested, is difficult to isolate due
to backgrounds, the dominant one being top-pair production.  
The backgrounds have been studied in
\cite{Tait:1999cf, backgrounds},
and kinematic cuts which can clearly isolate the $tW$ channel have been 
suggested there. A similar analysis including NLO effects has been
carried out in \cite{White:2009yt,Frixione:2008yi}. 
Recently, 
with integrated luminosities of 0.7 fb$^{-1}$ and 2.1 fb$^{-1}$ respectively,
ATLAS \cite{ATLAS0.7} and CMS \cite{CMS2.1}
 have presented results on a search for the $tW$ signal, where both $W$'s decay 
leptonically. 
They have been able to put an upper bound on the cross section.

In a recent work \cite{rs}, 
we examined in the context of the LHC 
the capability of the process c) above, viz., 
single-top production in
association with a $W$, for providing information on possible anomalous $tbW$
couplings in a model-independent way. The processes a) and b), while
sensitive to $tbW$ couplings, could also get contribution from other
processes like scalar exchange, and do not permit a model-independent
approach.  
In \cite{rs}, anomalous $tbW$ couplings were found to change the production
mechanism, giving rise to top polarization different from that predicted
by SM. In addition, top decay distributions can also get modified by
anomalous couplings. Thus, a study of decay distributions in the
laboratory (lab) frame were shown to
carry signatures of anomalous $tbW$ couplings firstly through the degree
of polarization, and secondly through the contribution to the decay
process itself.

 Tevatron provides the only existing direct limits on anomalous $tbW$
couplings through $W$ polarization measurements.  
These results \cite{WpolCDF} lead 
to a limit of 
0.3 at the 95\% confidence level (CL) on the magnitude of the tensor
coupling $f_{2R} $, the only
coupling relevant in high-energy processes. (The couplings are defined
later, in eqs. (\ref{tdecay}), (\ref{tbardecay})).
Early LHC data \cite{WpolLHC} on measurements of $W$ polarization in
$t\bar t$ production and $t$-channel single-top production allow a limit
of [$-0.6$, 0.3] to be put on $f_{2R}$ \cite{AguilarSaavedra:2011ct}.
There are
stringent indirect constraints coming from 
low-energy measurements of anomalous $tbW$ couplings. The measured rate
of $b\rightarrow s\gamma$ puts stringent 
constraints on the couplings $f_{1R}$ and $f_{2L}$ of about
$4\times10^{-3}$, since their contributions to $B$-meson decay get 
an enhancement factor of $m_t/m_b$
\cite{Cho:1993zb,Grzadkowski:2008mf}. 
The bound on the anomalous coupling $f_{2R}$, however, is weak, viz., 
[$-0.15$, 0.57] at
the  95 \% CL \cite{Grzadkowski:2008mf}. In \cite{Drobnak:2011wj} a
slightly more stringent bound has been found utilizing
$B_{d,s}-\bar{B}_{d,s}$ mixing. It is thus clear that better direct
limits on $f_{2R}$ are needed.

We propose making use of the same lab-frame variables which we employed
in \cite{rs} for $tW^-$ production, but for a study of CPV, we
would also need analogous variables for $\bar t W^+$ production. The
difference (in certain cases the sum) of these variables for the cases of 
$tW^-$ and $\bar t W^+$ production would be a measure of CPV.
For convenience, we will adopt a linear approximation
for the couplings. This enables determination of limits on the
CP-violating parameter which are independent of the parameter itself.
Moreover, we find that for maximal CPV, which we
define in terms of equality in magnitude of \t and \tbar couplings,
there is no quadratic contribution to the asymmetry, making the linear
approximation an excellent one.

At the LHC, the
initial state being $pp$ is not self-conjugate and it is not evident that 
a study of CP-conjugate final-state particles can reveal
the extent of CPV, if present. Nevertheless, 
associated $tW^-$ production is initiated by partons $b$ and $g$, whose respective
densities in $p$ are equal to those of the
conjugates $\bar b$ and $g$, from simple charge conjugation invariance
of strong interactions governing parton distributions in hadrons.

To see the consequences of CP invariance, consider a partial cross section for
$tW^-$ inclusive production at the LHC which may be written as
 \be\label{ppcross}
 \begin{array}{l}
 \!\!\!\!\!\!\!d\sigma \!\left(p(p_1) p(p_2)\! \to \!
 t(p_t,h_t)W^-(p_{W})X\!\right)\!\! = \!\! 
 \displaystyle\! \int\! \!dx_1dx_2 f_b (x_1) f_g (x_2) 
 d\hat \sigma_{bg\to tW^-}(x_1p_1,
 x_2p_2, p_t,h_t, p_{W}\!)
 \end{array}
 \ee
where $d\hat \sigma_{bg\to tW^-}$ is the corresponding parton-level
partial cross section, and $f_b$, $f_g$ are the densities of $b$, $g$
partons in the proton. $p_1$, $p_2$, $p_t$ and $p_{W}$ are
respectively
the momenta of the two protons, $t$ and $W^-$, and $h_t$ is the helicity
of $t$.  One can write an analogous expression for the partial cross
section $d\sigma \left(p(p_1) p(p_2) \to \bar t(p_{\bar t},h_{\bar
t})W^+(p_{W}) \bar X\right)$ for $\bar t W^+$ production.
CP invariance at the parton level, which implies
 \be
 d\hat \sigma_{bg\to tW^-}(x_1p_1, x_2p_2, p_t, h_t,p_{W}) 
 	= d\hat \sigma_{\bar bg\to \bar tW^+} (x_1p_1,          
 x_2p_2, p_{ t},-h_{ t}, p_{W}),   
 \ee
gives, for the hadron-level cross sections,
 \be
 d\sigma (pp \to t(p_t,h_t) W^-(p_{W})X) = 
 d\sigma (pp \to \bar t(p_{t},-h_{t}) W^+(p_{W})\bar
 X).
 \ee
A violation of this relation would signal CPV. Thus, CP invariance implies
equal and opposite longitudinal polarizations for $t$ and $\bar t$ in
the two processes.

\smallskip

\noindent
{\it 2. Effective vertices:}
We can write the most general 
effective vertices up to mass dimension 5 for the four distinct processes of 
\t decay, \tbar decay, \t production and \tbar production as 
 \bea\label{tdecay}
 V_{t\to bW^+} &=& - (g/\sqrt{2}) V_{tb} [ \gamma^\mu 
 	(f_{1L} P_L + f_{1R} P_R ) 
 	-i(\sigma^{\mu\nu}q_\nu/m_W)  (f_{2L} P_L + f_{2R}
 	P_R ) ],\\
 \label{tbardecay}
 V_{\bar t\to \bar bW^-} &=& - (g/\sqrt{2}) V_{tb}^*[\gamma^\mu 
 	(\bar f_{1L} P_L + \bar f_{1R} P_R ) 
 	-i(\sigma^{\mu\nu}q_\nu/m_W) (\bar f_{2L} P_L + \bar f_{2R}
 	P_R )],\\
 \label{tprod}
 V_{b^*\to tW^-} &=& - (g/\sqrt{2}) V_{tb}^* [ \gamma^\mu 
 	(f_{1L}^* P_L + f_{1R}^* P_R ) 
 	-i(\sigma^{\mu\nu}q_\nu/m_W) (f_{2L}^* P_R +
 	f_{2R}^* P_L ) ],\\
 \label{tbarprod}
 V_{\bar b^*\to \bar tW^+} &=& - (g/\sqrt{2}) V_{tb} [ \gamma^\mu 
 	(\bar f_{1L}^* P_L + \bar f_{1R}^* P_R ) 
 	-i(\sigma^{\mu\nu}q_\nu/m_W) (\bar f_{2L}^* P_R +
 	\bar f_{2R}^* P_L )],
 \eea
where $f_{1L}$, $f_{1R}$, $f_{2L}$, $f_{2R}$,
$\bar f_{1L}$, $\bar f_{1R}$, $\bar f_{2L}$, and $\bar f_{2R}$ are form
factors, $P_L$, $P_R$ are left-chiral and right-chiral projection
matrices, and $q$
represents the $W^+$ or $W^-$  momentum, as applicable in each case.
In the SM, at tree level, $f_{1L}=\bar f_{1L} = 1$, and all other form
factors vanish. New physics effects would result in deviations of
$f_{1L}$ and  $\bar f_{1L}$ from unity, and nonzero values for other
form factors.

At tree level, i.e., in the absence of any absorptive parts in the
relevant amplitudes giving rise to the form factors, the following
relations would  be obeyed: 
 \be\label{noabspart}
 \begin{array}{rccl}
 f_{1L}^* = \bar f_{1L};& f_{1R}^* = \bar f_{1R}; & f_{2L}^* = \bar
 f_{2R}; & f_{2R}^* = \bar f_{2L}.
 \end{array}
 \ee

This can be seen to follow from the fact that in the absence of
final-state interactions leading to absorptive parts, the effective
Lagrangian is hermitian. Thus, the hermiticity of the Lagrangian
 \be\label{lag}
 \begin{array}{rcl}
 {\mathcal L}_I & = & \displaystyle - (g/\sqrt{2}) \Big[V_{tb} \bar b \left\{
 	\gamma^\mu
         \left(f_{1L} P_L + f_{1R} P_R \right)
         -(\sigma^{\mu\nu}/m_W) \partial_\nu W^-_{\mu} 
 	\left(f_{2L} P_L + f_{2R}
         P_R \right) \right\} t \\
 &&\displaystyle + V_{tb}^* \bar t \left\{
         \gamma^\mu
         \left(\bar f_{1L} P_L + \bar f_{1R} P_R \right)
         -(\sigma^{\mu\nu}/m_W) \partial_\nu W^+_{\mu} 
         \left(\bar f_{2L} P_L + \bar f_{2R}
         P_R \right) \right\} b \Big] + {\mathrm H.c.},
 \end{array}
 \ee
from which the \t and \tbar decay and production amplitudes
(\ref{tdecay})-(\ref{tbarprod}) may be derived, implies the relations (\ref{noabspart}).

On the other hand, if CP is conserved, the relations obeyed by the form
factors are
\be\label{CP}
\begin{array}{rccl}
f_{1L} = \bar f_{1L};& f_{1R} = \bar f_{1R}; & f_{2L} = \bar
f_{2R}; & f_{2R} = \bar f_{2L}.
\end{array}
\ee

We can write the phases of the form factors as sums and 
differences of two phases, one corresponding to a nonzero absorptive
part, and another corresponding to CP nonconservation:
 \be\label{phases}
 \begin{array}{rl}
 f_{1L,R} = \vert f_{1L,R} \vert \exp{(i\alpha_{1L,R} + i\delta_{1L,R})};&
 \bar f_{1L,R} = \vert f_{1L,R} \vert \exp{(i\alpha_{1L,R} -
 i\delta_{1L,R})};\\
 f_{2L,R} = \vert f_{2L,R} \vert \exp{(i\alpha_{2L,R} + i\delta_{2L,R})};&
 \bar f_{2R,L} = \vert f_{2L,R} \vert \exp{(i\alpha_{2L,R} - i\delta_{2L,R})},\\
 \end{array}
 \ee
where, for convenience, magnitudes of the couplings are assumed to be
equal in appropriate pairs.
In case there are no absorptive parts ($\alpha_i = 0$), we get the
relations (\ref{noabspart}) as a special case. In case of CP
conservation, we get the relations (\ref{CP}) as a special case. In what
follows, we will deal with a more general case when both CPV
and absorptive parts are present. When calculating differences of
variables for $tW^-$ and $\bar tW^+$ production, we will encounter only
the combination
 \be\label{deltacp}
 \mathrm{\Delta_{2R}}\equiv ({\rm Re}\bar{f}_{2L}-\re f_{2R})
 = -2|f_{2R}|\sin\alpha_{2R}\sin\delta_{2R}.
 \ee
(We assume $f_{1L}=\bar f_{1L}=1$ and $V_{tb}=1$ throughout).
For $\rm \Delta_{2R}$ to be nonzero, both absorptive parts and CPV 
have to be present, as can be seen from eq. (\ref{deltacp}).

\begin{figure}[h]
\begin{center}
 \includegraphics[scale=1.0]{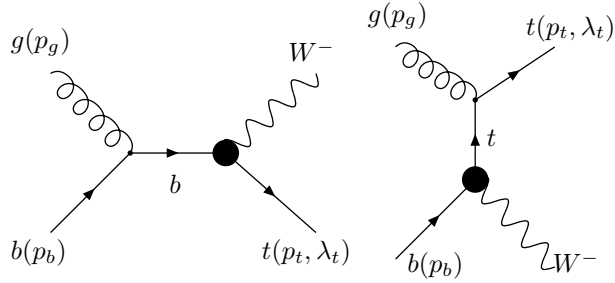}
\end{center}
\caption{ Feynman diagrams contributing to associated $tW^-$ production
at the LHC. The effective $tbW$ vertex is shown as a filled circle.}\label{feyngraph1}
\end{figure}

The effective vertices (\ref{tdecay})-(\ref{tbarprod})
contribute to the production and decay of the top quark and antiquark.
The parton-level Feynman diagram for $tW^-$ is shown in Fig.
\ref{feyngraph1}, with the effective vertex shown as a filled circle.
The effect of anomalous couplings would
show up in the production process in the total cross section, $t$ or
\tbar angular distributions and \t or \tbar polarizations. This is apart
from $W$ polarization, which we do not consider here. As for the decay
process, for a given production angle and polarization of $t$/\tbar, 
anomalous couplings would change the angular distribution of the decay
products.  

In what follows, we obtain the contribution of CP-violating $tbW$ couplings in
sums or differences of observables in $tW^-$ and $\bar tW^+$ production
and decay, and examine how well the couplings can be constrained by
them. 
We retain only $\re f_{2R}$ and $\re \bar f_{2L}$ , since these are the only ones which
contribute in the limit of vanishing bottom mass, which we set to zero. 
We will assume that the anomalous couplings are small and 
a linear approximation of the couplings found to be good for 
$|\re f_{2R}|$ up to about 0.05 \cite{rs}.

\smallskip 

\noindent
{\it 3. CP-violating asymmetries:}
We now formulate CP-violating asymmetries in the single-top production subprocess $gb\to tW^-$ and its 
CP-conjugated subprocess $g\bar{b}\to \bar{t}W^+$  
through CP-violating anomalous $tbW$ couplings.
In \cite{rs}, we have obtained analytical expressions for the spin density 
matrix for the top quark produced in the process $gb\to tW^-$ and for the top decay including the full contribution of anomalous $tbW$ 
couplings. The corresponding expressions for the CP-conjugated process $g\bar{b}\to \bar{t}W^+$ and $\bar{t}\to\bar{b}W^-$ can be obtained from 
those expressions by replacing $\re f_{2R}$ by $\re \bar{f}_{2L}$
taking appropriate helicity of the antitop.

The simplest asymmetry we consider is the rate asymmetry in the
production of $tW^-$ and $\bar{t}W^+$.
Assuming the equality of $b$ and $\bar b$ densities
in the proton, we can write, 
at linear order in the anomalous couplings, 
 \be
 \begin{array}{cc}
 \sigmat=\sigma_0+ \re f_{2R} \ \sigma_1,&
 \sigmatb=\sigma_0+ \re\bar{f}_{2L} \ \sigma_1 ,
 \end{array}
 \ee
where $\sigma_0$ is the SM cross section, which is identical for $pp\to
tW^-X$ as well as $pp \to \bar tW^+X$ and $\sigma_1$ is the cross
section arising 
from interference of the anomalous contribution with the SM amplitude,
for unit anomalous coupling. 
We can then write the fractional rate asymmetry, 
 \be\label{A:sigma-CPV}
 \begin{array}{ccc}
 \displaystyle
 {A}_\sigma^{CPV}=1/ (2\sigma_0)
 ( {\rm Re}\bar{f}_{2L}-{\rm Re}f_{2R}) \ \sigma_1
 &
 \displaystyle
 =
 1/ (2\sigma_0)\rm\Delta_{2R} \ \sigma_1.
 \end{array}
 \ee

In the linear approximation, the $t$ polarization $P_t$
and the $\bar t$ polarization $P_{\bar{t}}$ can be written as
 \bea
 P_t=P_t^0+\re f_{2R} \ P_t^1;~~~
 P_{\bar{t}}=-P_t^0-\re \bar{f}_{2L} \ P_t^1,
 \eea
where $P_t^0$ and $P_t^1$ are contributions from the SM and from the
pure anomalous couplings at linear order, respectively.
In the linear order in anomalous couplings, a  CP-violating asymmetry
$A_{P_t}^{CPV}$ in top and antitop polarization defined by 
$A_{P_t}^{CPV}=P_{\bar{t}}+P_t$ takes the form
 \bea
 A_{P_t}^{CPV}=-( \re\bar{f}_{2L}-\re f_{2R})P_t^1
 =-\rm\Delta_{2R}P_t^1.
 \eea

The asymmetries so far did not take into account top decay. A
measurement of these asymmetries would need full reconstruction of the
top. Thus the corresponding efficiencies would have to be taken into
account. We now look at asymmetries where an explicit decay channel of
\t or \tbar is taken into account.

A rate asymmetry may be found corresponding to the combined
production and semileptonic decay of the $t$ and $\bar t$, where both
the production and decay processes get contributions from the anomalous
couplings.
Using the narrow-width approximation, we can write the cross section for
the process $pp\to tW^-X\to b\ell^+\nu W^-$ as 
 \beq
 \sigma_{\ell^+W^-} \equiv
 \sigma(pp\to b\ell^+\nu W^-X)=\left(1/\Gamma_t\right)\times \left[\rho({pp\to
 tW^-X})\otimes\Gamma(t\to
 b\ell^+\nu)\right],
 \eeq
 where $\Gamma_t$ is total decay width of the top, $\rho({pp\to
tW^-X})$, $\Gamma(t\to
b\ell^+\nu)$ are respectively the production and decay density matrices 
(with appropriate integration of the phase space carried out),
and 
$\otimes$ denotes a matrix product.
The density matrix formalism is used here to ensure proper spin
coherence between production and decay.
Writing an analogous expression for the cross section
$\sigma_{\ell^-W^+}$
for the conjugate process,
we define the asymmetry in charged-lepton production rates as
 \beq
 A_\ell^{CPV}= [\sigma_{\ell^+W^-} - \sigma_{\ell^-W^+}]/
 [\sigma_{\ell^+W^-} + \sigma_{\ell^-W^+}].
 \eeq
  
$A_\ell^{CPV}$ gets contribution not only from the asymmetry in \t/\tbar
production, but
also from the anomalous coupling in the decay into the leptonic channel.

In all the following cases, we write the generic asymmetry in the $tW^-$
process as 
as $A^-$, and that in $\bar tW^+$ production as $A^+$. We 
can then write, in the linear approximation,
 \bea
 A^- = A^0 + {\rm Re} f_{2R}\, A^1;~~~
 A^+ = A^0 + {\rm Re} \bar f_{2L}\, A^1,
 \eea
where $A^0$ is the asymmetry in the SM, and $A^1$ is the contribution
coming from the interference between the term with 
anomalous couplings and the SM term, for unit value of the coupling.
The CP-violating asymmetry is then
 \be
 A^{CPV} \equiv A^+ - A^- = {\rm \Delta_{2R}}\,A^1.
 \ee

The azimuthal distributions of the charged lepton 
and the $b$ quark arising from the top decay are
sensitive to anomalous couplings \cite{rs}
and we define the corresponding azimuthal asymmetries as
 \beq
 A_{\phi_{\ell,b}}=[\sigma(\cos\phi_{\ell,b}>0)-\sigma(\cos\phi_{\ell,b}<0)]/
 [\sigma(\cos\phi_{\ell,b}>0)+\sigma(\cos\phi_{\ell,b}<0)],
 \eeq
where $\phi_\ell$, $\phi_b$ are respectively the azimuthal angles of the charged
lepton and the $b$ quark w.r.t. the
top production plane. An analogous asymmetry can be defined for the
charged lepton and $b$ arising in antitop production. 
The CP-violating asymmetries $A^{CPV}_{{\phi_\ell}}$ and
$A^{CPV}_{{\phi_b}}$ are then defined as the
differences between the relevant 
azimuthal asymmetries from the $tW^-$ process and
the conjugate process.

The charged-lepton energy distribution
is also sensitive to  anomalous $tbW$ couplings \cite{rs}, and  
the distributions for SM and different anomalous couplings
all intersect at about $E_\ell^C=62$ GeV. To quantify the differences in
distributions for different couplings, we constructed in \cite{rs} an 
asymmetry around the intersection energy $E_\ell^C$ of the curves, defined by
 \begin{equation}
  A_{E_\ell}=[\sigma(E_\ell
 <E_\ell^C)-\sigma(E_\ell>E_\ell^C)]/[\sigma(E_\ell <
 E_\ell^C)+\sigma(E_\ell > E_\ell^C)],
 \label{Enasy}
 \end{equation}
Using this, and the corresponding asymmetry for antitop production and
decay, 
we define a CP-violating asymmetry
$A_{E_\ell}^{CPV}$ as their difference.

\begin{figure}[h!]
\begin{center}
  \includegraphics[angle=270,width=3.15in]{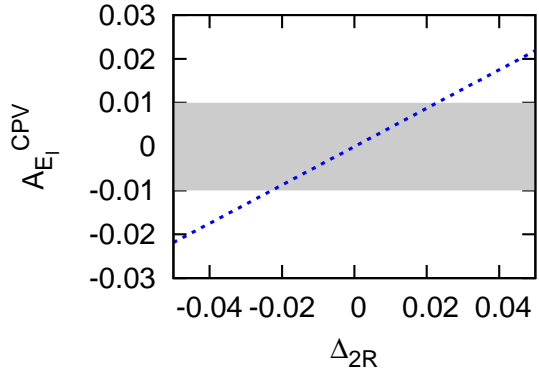}
\caption{\sl The CP-violating energy asymmetry of charged leptons from
$t$ and \tbar decay 
 as a function of ${\mathrm \Delta_{2R}}$ for cm energy 14 TeV. The band 
corresponds to the 1$\sigma$ error interval in the SM for 
integrated luminosity 10 fb$^{-1}$.} 
\label{Asym-Energy}
\end{center}
\end{figure}

\smallskip

\noindent 
{\it 3. Numerical Results:} We now evaluate the various asymmetries defined.
For our numerical analysis, we consider for the LHC a cm 
energy $\sqrt{s}=7$ TeV  for integrated luminosity 
$L=5$ fb$^{-1}$, and  $\sqrt{s}=14$ TeV with 
$L=10$ fb$^{-1}$. While the present run of the LHC is at $\sqrt{s}=8$
TeV, we have checked that our results for this energy would be
similar to those for $\sqrt{s}=7$ TeV.
We use the leading-order parton distribution
function sets of CTEQ6L \cite{cteq6} with a factorization scale of $m_t=172.6$ GeV. 
Values of parameters used are: $M_W=80.403$ GeV, 
$\alpha_{em}(m_Z)=1/128$ and $\sin^2\theta_W=0.23$.

We have imposed acceptance cuts on 
the charged-lepton rapidity and transverse momentum, viz., $|\eta|<2.5$, 
and $p_{T}^\ell>20$ GeV. We have assumed an ideal situation of 100\%
efficiency for $W$ and $b$-jet identification, since we have used
analytical expressions integrated over the full kinematic range for $W$
and $b$. We later discuss the effect of realistic cuts needed for
isolating the $Wt$ final state.
In order to discriminate the charge of the lepton which comes from top
decay, 
we thus restrict ourselves to leptonic decay of the top
and hadronic decay of the $W$.

In the linear
approximation, the asymmetry is given by the parameter $\rm \Delta_{2R}$
multiplied by the quantity $A^1$ for each asymmetry
which is given in Table \ref{combined} against the
corresponding asymmetry. 
As can be seen, 
the lepton energy asymmetry is the largest among the asymmetries
constructed out of decay distributions.
Also shown in Table 1, for various asymmetries, are the possible 
$1\sigma$ limits on the
CP-violating parameter $\rm \Delta_{2R}$, using for the
statistical uncertainty in $A^{CPV}$ 
the value $\delta A^{CPV} = \sqrt{2}/\sqrt{L\sigma_{SM}}$,
where $\sigma_{SM}$ is the SM cross section for the relevant final state.
For this, we have taken only one of the leptonic decay modes for the
top, while including all hadronic channels for the decay of the $W$.
The results are shown in Table 1.
\begin{table}
\centering
 \begin{tabular}{|c|c|c|c|c|c|c|c|}
 \hline
 $\sqrt{s}$ &      & $A_{\sigma}^{CPV}$ & $A_{P_t}^{CPV}$ &
 $A_\ell^{CPV}$ &
 $A_{\phi_\ell}^{CPV}$  & $A_{E_\ell}^{CPV}$ & $A_{\phi_b}^{CPV}$ \\
 \hline
   & $A^{CPV}$ for $\rm \Delta_{2R}=1$ & 0.508 & 0.526 & 0.362 & 0.198 &
 0.464 & 0.093\\
 7 TeV  
   & Limit ($L=5~{\rm fb}^{-1}$) &   
 $0.0183$ &$0.0180$ &$0.106$ & $0.175$ & $0.075$& 0.127 \\
 \hline
 & $A^{CPV}$ for $\rm \Delta_{2R}=1$ & 0.502 & 0.536 & 0.369 & 0.186 &
 0.438 & 0.086 \\
 14 TeV  & Limit ($L=10~{\rm fb}^{-1}$) 
 & $0.00542$ &$0.00506$ & $0.0298$& $0.0343$ & 0.0228 & 0.0386\\
 \hline
 \end{tabular}\label{combined}
\caption{\sl The value of various CP-violating asymmetries for unit
value of the  
CP-violating combination of couplings, $\rm \Delta_{2R}$, 
and the corresponding $1\sigma$ limits on $\vert \rm \Delta_{2R}\vert $
possible
at the LHC.}
\end{table}
For the estimation of limits using $A_{\sigma}^{CPV}$ and $A_{P_t}^{CPV}$, we assume
100\% efficiency in the detection of the top, as well in the measurement
of its polarization. The corresponding idealized limits are for
comparison with the remaining more realistic ones obtained from kinematics of
decay products.

We see from the Table 1 that $1\sigma$ limits on $\vert \rm \Delta_{2R}
\vert$ possible for $\roots=7$ TeV with 
$L=5$ fb$^{-1}$ are of the order of about $0.1$, perhaps at the limit of
validity of our linear approximation. The limit from $A_{P_t}^{CPV}$ is
nominally better, but cannot be realized since the polarization cannot
be measured with 100\% accuracy.
For $\roots = 14$ TeV, on the other hand, the limits are of
the order of a few times $10^{-2}$, well within the validity range of
the approximation. 
In Fig. \ref{Asym-Energy}  we show $A_{E_\ell}^{CPV}$, which
gives the best limit on $\rm \Delta_{2R}$, as a function 
of $\rm \Delta_{2R}$. 
Also shown in the figure is 
the $1\sigma$ band corresponding to the statistical uncertainty.
The limit on $\vert \rm \Delta_{2R}\vert$ which would come from  
the measurement of $A_{E_\ell}^{CPV}$ is $2.28\times 10^{-2}$.

We now discuss how the results of Table 1 would be affected on 
inclusion of realistic cuts needed for the isolation of the $tW$ events
and suppression of background. For this we make use of the analysis
carried out recently by the ATLAS and CMS collaborations for the LHC
data. 

We determine the factor by which the theoretical number for the Wt events 
get reduced by cuts in a single leptonic
channel in top as well as W decay, making use of
Table 2 of \cite{ATLAS0.7} and also Table 2 of \cite{CMS2.1}.
We find this factor to be approximately 0.1. 
These searches use leptonic decay channels of W 
whereas a
search for CPV should necessarily use the two-jet decays of W.
For such a final state, ref. \cite{backgrounds} indicates a somewhat 
better efficiency
than 0.1 for each channel, so we are being somewhat conservative. 

We then assume that the same factor can be used in our CP-asymmetry
analysis (also at 14 TeV) to get an idea of the number 
of events which survive the cuts. 
Undoing the leptonic cut already used in Table 1, and including all
leptonic channels (we assume detection efficiency of 0.6 for $\tau$'s),
we arrive at a correction factor of 0.325 for the event rates, and 1.754
for the limits in Table 1.

We thus see that in the best scenario of $\sqrt{s}=14$ TeV with $L=10$
fb$^{-1}$, a realistic limit of about 0.04 on $\vert \rm \Delta_{2R}\vert$
should be possible using the leptonic energy asymmetry.

\smallskip 

\noindent
{\it 4. Conclusions and Discussion:}
We have investigated the possibility of measuring CP-violating $tbW$
couplings at the LHC through the conjugate processes of $tW^-$ and $\bar
tW^+$ production. We proposed a number of CP-violating asymmetries which
would be sensitive to the CP-odd combination of couplings $\mathrm{
\Delta_{2R}} \equiv \bar f_{2L} - f_{2R}$, the difference of the 
couplings associated with the top and the antitop.

We conclude from our analysis that the $tW$ mode of single-top production is  
a good alternative process to look for CP-violating $tbW$ couplings apart from 
a comparison of $t$ and $\bar t$ decays in top-pair production. We find that the energy asymmetry, 
$A_{E_\ell}^{CPV}$, is the most sensitive to the CP-violating parameter
and would enable a limit of about $0.04$ to
be placed on $\rm\vert\Delta_{2R}\vert$, for a cm energy of 14 TeV and
integrated luminosity of 10 fb$^{-1}$. The limits possible for an
operational energy of 7 TeV for integrated luminosity of up to 5
fb$^{-1}$ are not as good, being at the level of 0.1-0.2.

Higher-order QCD and electroweak effects can give
rise to a partonic-level forward-backward asymmetry at the per cent
level in the background process of $t\bar t$
production.
This asymmetry can, because of valence-sea PDF difference, induce 
in the lab frame observables 
a charge asymmetry which does not originate in CPV.
However, this is present only in the sub-dominant quark-antiquark
process. Moreover, how our leptonic observables would be affected
needs detailed investigation, beyond the scope of this work.
Table 1 shows that $A_{E_l}^{CPV}$ is
fairly close in magnitude to $A_{P_t}^{CPV}$, and
seems to be a good measure of the polarization asymmetry. Since
polarization can only arise from chiral couplings,
higher-order QCD effects cannot generate it. $A_{E_l}^{CPV}$,
at least, is therefore unlikely to get significant contribution from
higher-order QCD effects. Top polarization due to electroweak
effects is tiny \cite{godbole}.

All in all we conclude that it would be possible to obtain limits on the
CP-violating anomalous coupling in the region of $10^{-2}$ by employing
the energy asymmetry of the charged lepton. Since this is the first
investigation in this direction, these numbers     
corresponding to direct limits should be taken as encouraging. 
It would be worthwhile carrying out a more detailed and refined analysis. 

Various extensions of the SM would have specific predictions 
for these anomalous couplings. For example, the contributions to 
these form factors in 2HDM, 
MSSM and top-color assisted 
Technicolor model have been evaluated in 
\cite{Bernreuther:2008us}. The predictions in these models, especially
 for the imaginary parts of
the couplings, seem to be much smaller than
the limits we expect from LHC with $L=10$ fb$^{-1}$.

 \smallskip
 \noindent Acknowledgment: SDR acknowledges financial support from the Department
 of Science and Technology, India, under the J.C. Bose National
 Fellowship programme, grant no. SR/SB/JCB-42/2009.

\end{document}